\documentclass[5p,10pt]{elsarticle}
\usepackage{xfrac}
\usepackage{color}
\usepackage{multirow,bigdelim}
\usepackage{varwidth}
\usepackage{setspace}
\usepackage{calc} 
\usepackage{graphicx}
\usepackage{amssymb}
\usepackage{amsmath}
\usepackage{mathtools}
\usepackage{dcolumn}
\usepackage{booktabs}
\usepackage{colortbl}
\usepackage{rotating}
\usepackage{sidecap}
\usepackage{subfig}
\usepackage{changes}
\usepackage{mathtools, cuted}
\newcommand{\bigzero}{\mbox{\normalfont\Large\bfseries 0}}
\newcommand{\rvline}{\hspace*{-\arraycolsep}\vline\hspace*{-\arraycolsep}}

\usepackage{hyperref}

\makeatletter
\def\@author#1{\g@addto@macro\elsauthors{\normalsize%
    \def\baselinestretch{1}%
    \upshape\authorsep#1\unskip\textsuperscript{%
      \ifx\@fnmark\@empty\else\unskip\sep\@fnmark\let\sep=,\fi
      \ifx\@corref\@empty\else\unskip\sep\@corref\let\sep=,\fi
      }%
    \def\authorsep{\unskip,\space}%
    \global\let\@fnmark\@empty
    \global\let\@corref\@empty  
    \global\let\sep\@empty}%
    \@eadauthor={#1}
}
\makeatother

\newcommand{\atomicarrow}{\mathrlap{\bigcirc}{\leftarrow}}

\newcommand{\orderof}[1]{\mathcal{O}\left(#1\right)}

\newcommand{\qop}[1]{\hat{#1}}
\newcommand{\mat}[1]{{\bf{#1}}}

\biboptions{comma,square,sort&compress}

\newcounter{bla}

\begin{document}

\begin{frontmatter}

\date{February 2023}

\title{The N-shaped partition method: A novel parallel implementation of the Crank Nicolson algorithm}

\begin{abstract}
We develop an algorithm to solve tridiagonal systems of linear equations, which appear in implicit finite-difference schemes of partial differential equations (PDEs), being the time-dependent Schr\"{o}dinger equation (TDSE) an ideal candidate to benefit from it. Our N-shaped partition method optimizes the implementation of the numerical calculation on parallel architectures, without memory size constraints. Specifically, we discuss the realization of our method on graphics processing units (GPUs) and the Message Passing Interface (MPI). 
In GPU implementations, our scheme is particularly advantageous for systems whose size exceeds the global memory of a single processor.
Moreover, because of its lack of memory constraints and the generality of the algorithm, it is well-suited for mixed architectures, typically available in large high performance computing (HPC) centers. We also provide an analytical estimation of the optimal parameters to implement our algorithm, and test numerically the suitability of our formula in a GPU implementation. Our method will be helpful to tackle problems which require large spatial grids for which ab-initio studies might be otherwise prohibitive both because of large shared-memory requirements and computation times.

\end{abstract}

\author{Y.~Lutsyshyn} 
\author{F.~Navarrete\corref{cor3}}
\cortext[cor3]{Corresponding author.}
\ead{francisco.navarrete@uni-rostock.de}
\author{D.~Bauer}
\ead[url]{www.physik.uni-rostock.de/qtmps}
\address{Institute of Physics, University of Rostock, 18051 Rostock, Germany}


\end{frontmatter}

\section{Introduction}

The necessity to solve tridiagonal systems of linear equations arises naturally in many scientific and engineering studies such as
spectral Poisson solvers \cite{hockney1965acm}, fluid simulations \cite{kass1990cgit}, can be used in many-body perturbation theory calculations \cite{rubio2002rmp}, as well as in implicit 
finite-difference sche\-mes for the solution of partial differential equations (PDEs), such as the time-dependent Schr\"{o}dinger equation (TDSE). Among the different methods to solve the TDSE, which have different scaling and accuracy, we can mention the spectral method \cite{feit1982jcp}, the leap-frog method \cite{colgan2002pra, hu2004jpb}, the combination of the finite-element discrete variable representation (FEDVR) with the real-space product (RSP) algorithm\cite{schneider2006pre} and the Crank-Nicolson method \cite{mccullough1971jcp}. Our work is based on the implementation of the latter technique, for which the solution of tridiagonal matrices is the most important computational task. The diagonal band comes from the finite-difference approximation of the differential operators. The Thomas elimination method allows solving such systems efficiently. It scales as the first order of the matrix size, requires a small number of arithmetic operations, and has sequential memory access. It is stable and resilient to the accumulation of numerical errors. Even though the speed-up of many numerical algorithms can be greatly improved by their implementation in parallel architectures, unfortunately, the Thomas method is serial by essence. 
The need to parallelize the tridiagonal problem was recognized early on, with many parallel solvers proposed over the years. For a review, see \cite{BurrageBook1995}.
The parallel methods include recursive doubling 
\cite{Stone1975-ParallelTridiagonalEquationSolvers,%
Laub1989-ARecursiveDoublingAlgorithmForSolutionOfTridiagonalSystemsOnHypercubeMultiprocessors},
cyclic (even-odd) reduction 
\cite{LambiotteVoigt1975-TheSolutionOfTridiagonalLinearSystemsOnTheCDCSTAR100Computer,Heller1976-SomeAspectsOfTheCyclicReductionAlgorithmForBlockTridiagonalLinearSystems} (which requires 
a synchronization after every reduction stage
\cite{Johnsson1987-SolvingTridiagonalSystemsOnEnsembleArchitectures}),
domain decomposition methods and the partition methods.
The latter three approaches, including hybrid methods  
\cite{%
Zhang2010-FastTridiagonalSolversOnTheGPU,%
Davidson2011,%
Hwu2011-AScalableTridiagonalSolverForGPUs}
continue to be actively used. In addition, we have to mention that this problem could as well be solved in terms of the Woodbury formula\cite{numrec}, by
rewriting a big tridiagonal matrix as the sum of a block diagonal matrix (composed 
of tridiagonal blocks) plus a correction. A comparison with this method is performed in both methodology and complexity at the end of this work.

Many applications in the last decade have been specifically tuned to GPU:
\cite{%
Zhang2010-FastTridiagonalSolversOnTheGPU,%
Davidson2011,%
Boo2012,%
Doallo2015-NewTridiagonalSystemsSolversOnGPUArchitectures,%
Hwu2011-AScalableTridiagonalSolverForGPUs}. For this type of architecture one of the main challenges is solving matrices which do not fit in the shared memory. Even though recent works \cite{Davidson2011,doallo2018ieee}, based on the Wang and Mou \cite{wangmou1991ieee} Divide-and-Conquer strategy circumvent this constraint, they still face limitations when the problem size is bigger than the global memory of a single GPU and the work has to be distributed among many of them. An overview of the GPU implementations can be found in \cite{incollection}.

Amodio and Brugnano \cite{Amodio1992} developed a unifying parallel factorization approach for the tridiagonal problem and considered the benefits of the resulting methods. They also considered the application of the methods based on the hardware that was then available
\cite{Amodio1993}. 
The method that we consider here is a factorization method that falls within their framework.
The nature of parallel computer hardware has undergone considerable changes since the parallel tridiagonal solvers were originally developed. Here we revisit the benefits of the parallelization scheme from the viewpoint of modern massively parallel systems. We highlight the additional benefits of our novel factorization approach. In addition to the usual benefits of the partition method, we recognize that this approach includes the cache-beneficial access to the memory on individual processes; a non-blocking communication stage; integration with the forward step of the Crank-Nicolson method; integration between the substitution and reduction stages of the successive implicit steps; integration between successive iterations of the tridiagonal solver; and, importantly, the ability to mask the communication stage by computation. Some of these benefits become apparent only in conjunction with the particular PDE and the method for solving it. For this reason, details of the numerical problems that are of interest to us are also outlined. We also consider in some detail a recursive application of our algorithm.

This work is organized as follows. In Sec.~\ref{sec:cntdse} we outline
the Crank-Nicolson method for the solution of the 1D TDSE. 
In Sec.~\ref{sec:need_parallel} we review the need to parallelize the latter algorithm. We continue in Sec.~\ref{sec:method} with a full description of our novel N-shaped partition method. We describe in Sec.~\ref{sec:performance} the performance of the method regarding memory handling, communication cost and efficiency. Next, in Sec.~\ref{sec:implementation} we describe the GPU and the MPI implementation of our method. Finally we provide a brief summary and our conclusions in Sec.~\ref{sec:conclusion}. Unless specified otherwise we use atomic units in this work when discussing its application in physics. 

\section{The Crank-Nicolson method for the solution of the TDSE \label{sec:cntdse}}

As we discussed in the previous section, the method we developed can be used to 
solve the TDSE in one spatial dimension (1D), which can be written as
\begin{equation}
i\partial_t \psi(t) = \hat{H}(t) \psi(x,\,t)\;,
\label{eq:PDE}
\end{equation}
where $\psi$ is the wave function, and the operator $\qop{H}$
is the Hamiltonian of the system under consideration.

It is a standard procedure to solve Eq.~(\ref{eq:PDE}) using the Crank-Nicolson method  \cite{CrankNicolson1947,CrankNicolson1996} which
approximates the time propagator of the TDSE as
\begin{equation*}
\exp\left(-i\int_t^{t+\Delta t}\qop{H}(t')dt'\right) 
= \frac{1-\frac{i\Delta t}{2} \qop{H}(t+\Delta t/2)}
       {1+\frac{i\Delta t}{2} \qop{H}(t+\Delta t/2)} +\mathcal{O}(\Delta t ^ 3)\;.
\end{equation*}
where $ \Delta t$ is the time step.
The time propagation of the wave function is then obtained by solving
\begin{equation}
\qop{T} \psi(x,\,t+\Delta t) = \qop{T}^\dagger\psi(x,\,t) \;,
\label{eq:Cn_both_steps}
\end{equation}
with 
\begin{equation*}
\qop{T}=1+ \frac{i\Delta t}{2}\,\qop{H}\!\left(t+\frac{\Delta t}{2}\right)\;,
\end{equation*}
for $\psi(x,t+\Delta t)$. The method can be seen as equating the results of an explicit forward half-step $\qop{T}^\dagger \psi(x,\,t)$
and an implicit backward half-step $\qop{T} \psi(x,\,t+\Delta t)$. 

For solving this problem numerically we need to represent the functions and
operators on a computational grid. For instance, in position space, we may discretize the differential operators in $\qop{H}$ by finite-difference approximations, which
translates Eq.~(\ref{eq:Cn_both_steps}) into a matrix equation with tridiagonal matrices $\mat{T}$ and $\mat{T}^\dagger$. Each time propagation step involves the implementation of an explicit forward half-step
\begin{equation}
\mat{d}=\mat{T}^\dagger \mat{\Psi}(t)\;,
\label{eq:ForwardStep}
\end{equation}
and an implicit backward half-step,
\begin{equation}
\mat{T} \mat{\Psi}(t+\Delta t)=\mat{d}\;.
\label{eq:BackwardStep}
\end{equation}
Propagating for a single time step requires solving Eq.~\ref{eq:BackwardStep} for $\mat{\Psi}(t+\Delta t)$.
The above outlines the Crank-Nicolson method as it is used for solving the 1D TDSE. The method provides global accuracy to the second order in $\Delta t$,
and it is unconditionally stable. 

Even though for a system with $n_{tot}$ spatial grid points inverting the corresponding matrices might seem to imply performing $\orderof{n_{tot}^2}$ operations, 
the tridiagonal matrix Eq.~(\ref{eq:BackwardStep}) can be solved 
in just $\orderof{n_{tot}}$ by the Thomas algorithm. 
It requires $8n_{tot}$ arithmetic operations, which can be written as $2n_{tot}$ computationally expensive divisions, and $3n_{tot}$ computationally cheap multiplications and additions.
This computational cost is small enough to be comparable with the memory bandwidth. The Thomas method requires only consecutive memory access, allowing a highly efficient use of modern cacheing memory architectures. Similar considerations apply as well to the explicit step.

\section {The need to parallelize the Crank-Nicolson method} \label{sec:need_parallel}
The most straightforward way to implement the Crank-Nicolson method is by means of a serial algorithm. This algorithm is broadly used, for instance, in the simulation of the interaction of intense lasers with atomic and molecular targets. Even though this standard way to implement it has been successfully applied for different targets and laser parameters, for some applications, parallelizing its implementation might be very advantageous, and for others it might be the only option. For instance, in 1D, the interest to simulate
the interaction of matter with mid-IR intense laser pulses
requires substantially large numerical grids together with a need to reduce
the time step and to increase the simulation time. Combined, these requirements lead to demanding calculations, often taking several days or even weeks when implementing the algorithm serially. Furthermore, the number of spatial grid points might make the computation prohibitive due to the memory requirements for allocating the vectors necessary for the serial calculation. Here we will focus on solving 1D systems.

Solving the TDSE beyond 1D is frequently a demanding computational task. 
In higher dimensions, one can employ the operator splitting 
and thus reduce a single propagation step to solve a series of one-dimensional problems of the form (\ref{eq:Cn_both_steps}). It may seem that such a case may be trivially parallelizable. However, this is the case only for shared memory systems with uniform memory access. Large systems with distributed memory in fact cannot take advantage of this approach as the costs of moving the data around the network by far outweighs the costs of the numerical calculation. Thus a viable parallelization of a higher dimensional system must maintain memory locality. Each parallel process should only have to work with its own part of the function. While the method discussed here applies as-is to 1D systems, we will discuss the prospects of extending the method in Section \ref{sec:nd_considerations}.

In three dimensions, dynamics of an atom in strong radiation fields can be efficiently solved by using an expansion in spherical harmonics, such as proposed by Müller \cite{muller1999lphys} and implemented in the Qprop package
 \cite{Qprop,Qprop2,Qprop3}. 
In this case, one finds several effectively 1D equations of the form of Eq.~(\ref{eq:Cn_both_steps}) for the radial components of the wave function, with a coupling in the angular momentum numbers.
This is also true in, for instance, laser-nanoparticle interaction, where the target can be tens or hundreds of nanometers of diameter. 
Here one will also benefit directly from an efficient parallelization of the one-dimensional problem as described in our work.

\section{The N-shaped partition method \label{sec:method}}

In this section we describe our novel algorithm to parallelize the solution of a set of linear equations given by Eq.~(\ref{eq:BackwardStep}), i.e. the implicit backward half-step,
where $\mat{\Psi}$ is the vector of unknowns, and $\mat{T}$ is a tridiagonal matrix, both of size $n_{tot}$. We label their elements with indices running from $1$ to $n_{tot}$.
Conventionally, we label the diagonals of $\mat{T}$ as vectors $\mat{b}$, $\mat{a}$ and $\mat{c}$.
That is, $b_i\equiv T_{i,i-1}$, $a_i\equiv T_{i,i}$, $c_i\equiv T_{i,i+1}$, and
Eq.~(\ref{eq:BackwardStep}) represents a set of equations of the form
\begin{equation}
b_i \psi_{i-1} + a_i \psi_i  + c_i \psi_{i+1} = d_i \;,
\label{eq:algorithm}
\end{equation}
where all elements can be complex-valued. By construction, $b_1=0$ and $c_{n_{tot}}=0$ and thus we assume that $\psi_0=\psi_{n_{tot+1}}=0$ (reflecting boundary conditions). 
From Eq.~(\ref{eq:algorithm}), we can see that the algorithm requires memory to store the $5n_{tot}$ elements of vectors $\mat{a},\,\mat{b},\,\mat{c},\,\mat{d}$, and $\mat{\Psi}$.
No substantial additional memory is required.

\subsection{Structure of the algorithm}
\begin{figure*}
\begin{centering}
\def\arraystretch{2.5}
\begin{tabular}{|c|c|c|}  \hline
\includegraphics[width=.27\textwidth]{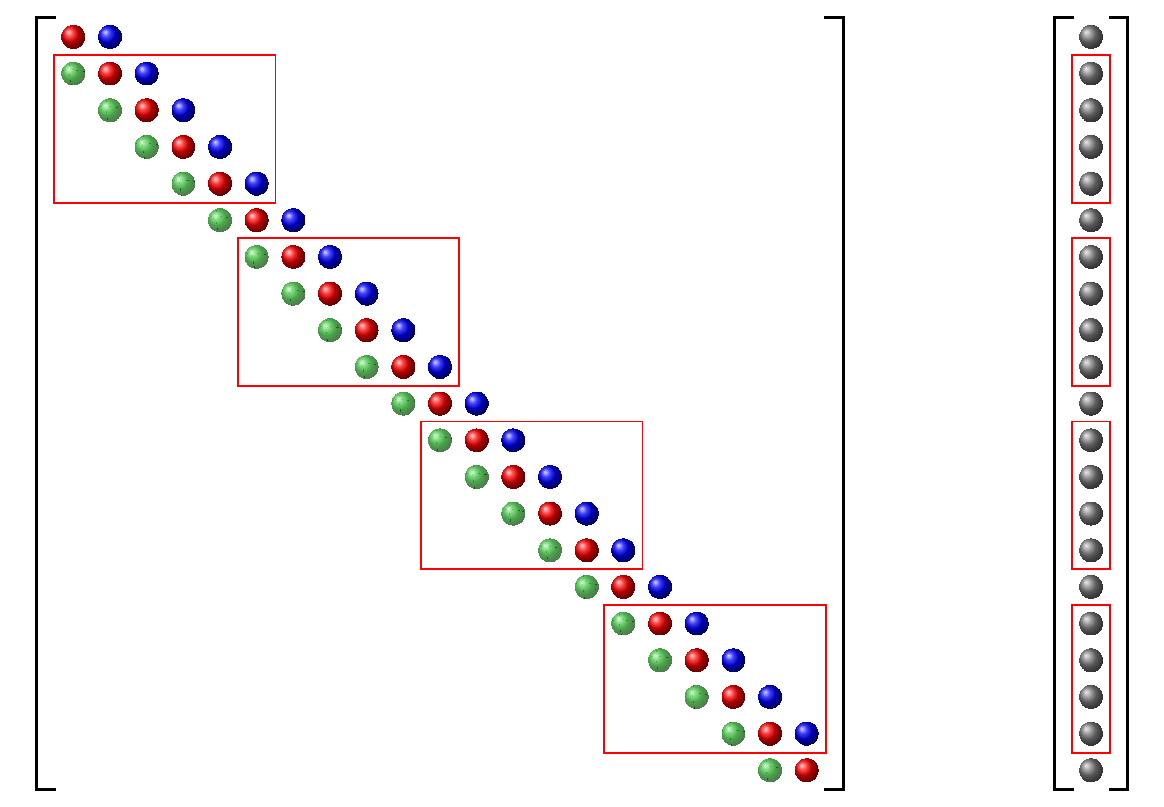} (a) & 
\includegraphics[width=.27\textwidth]{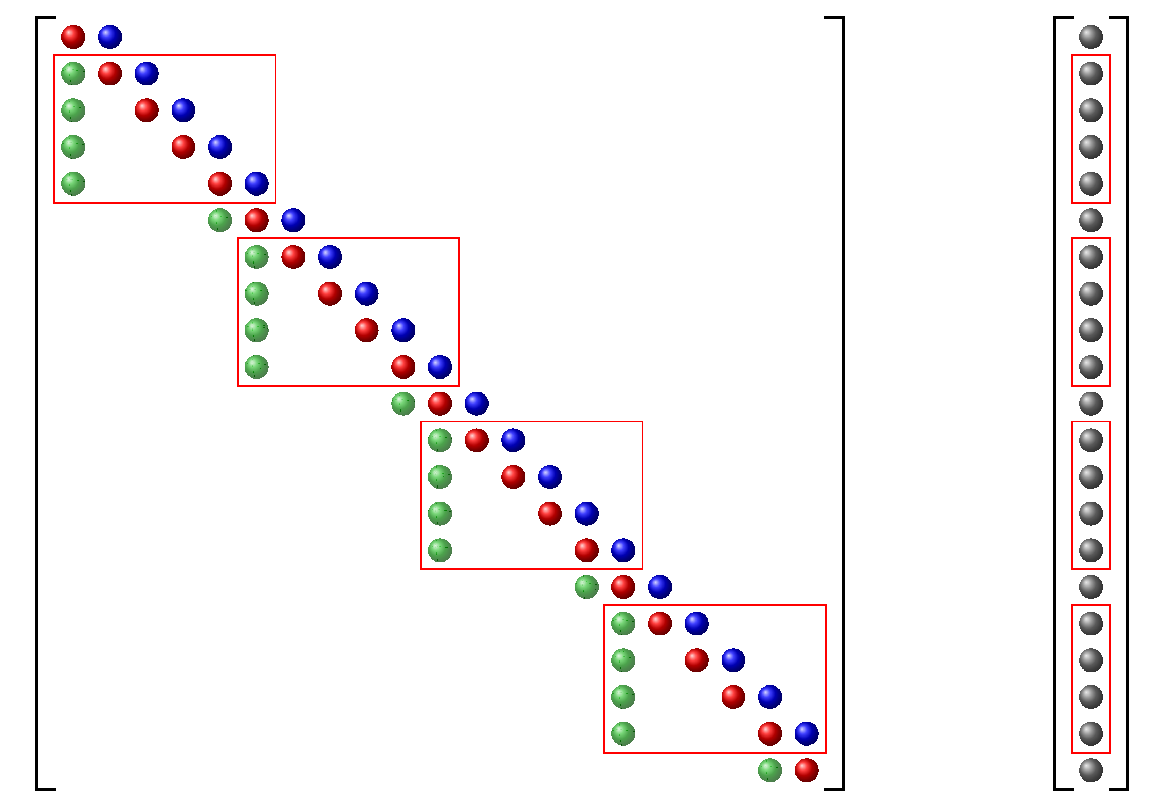} (b) & 
\includegraphics[width=.27\textwidth]{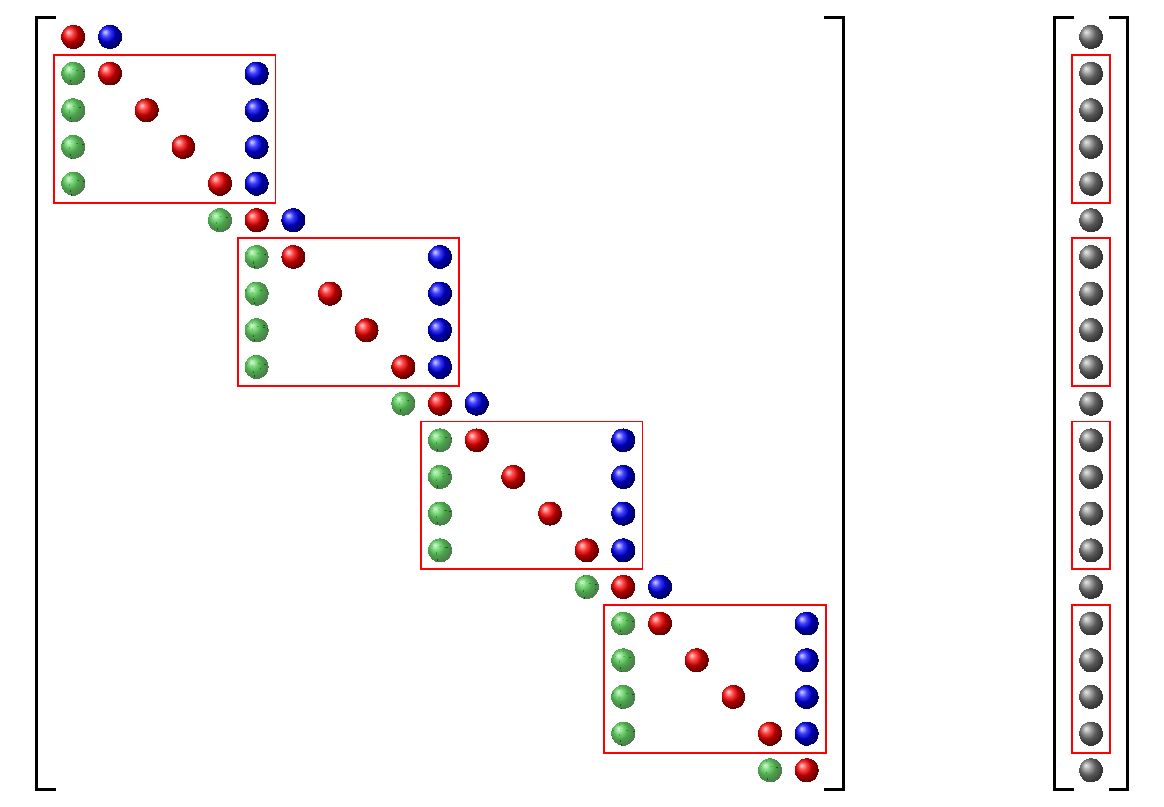} (c) \\ \hline
\includegraphics[width=.27\textwidth]{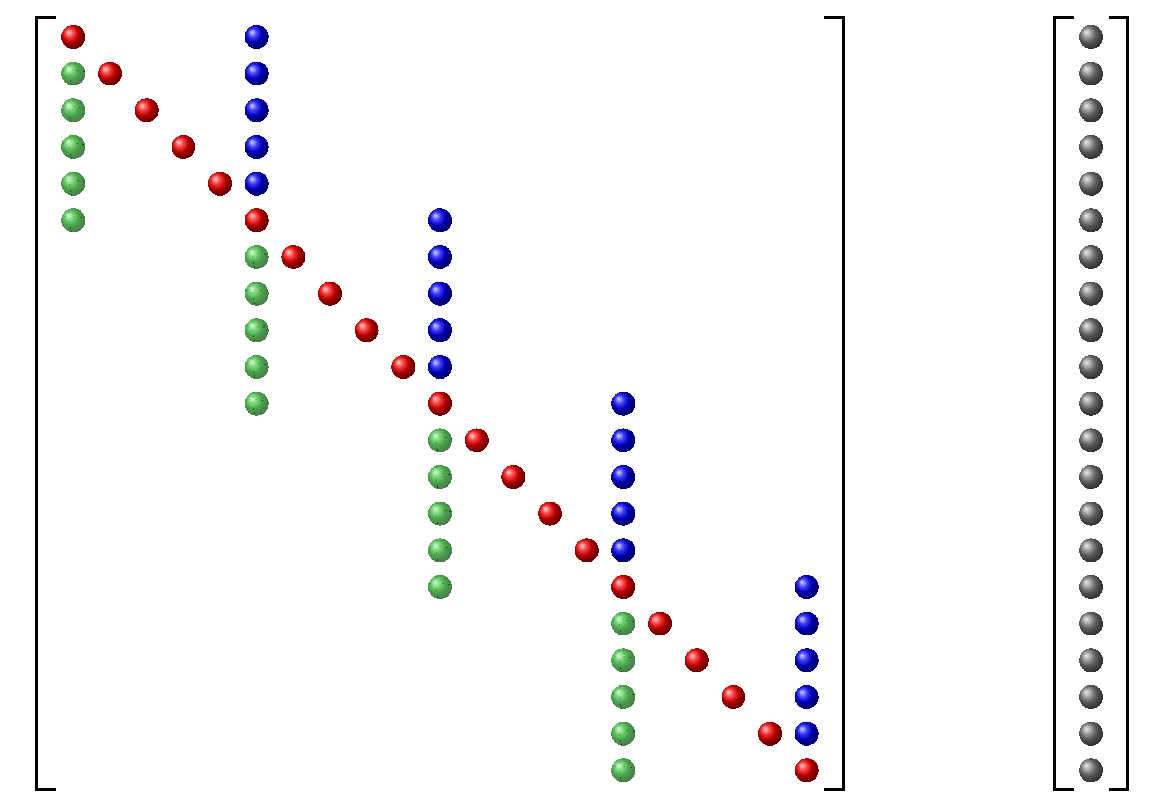} (d) & 
\includegraphics[width=.27\textwidth]{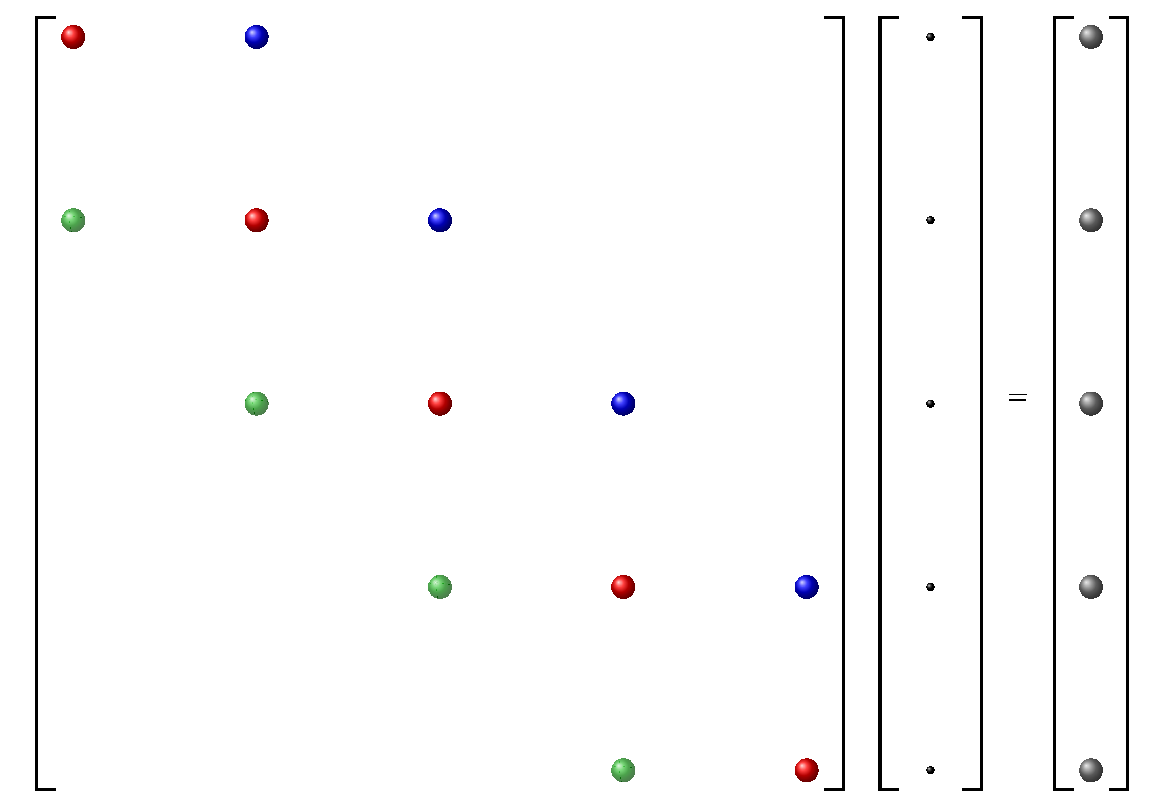} (e) & 
\includegraphics[width=.27\textwidth]{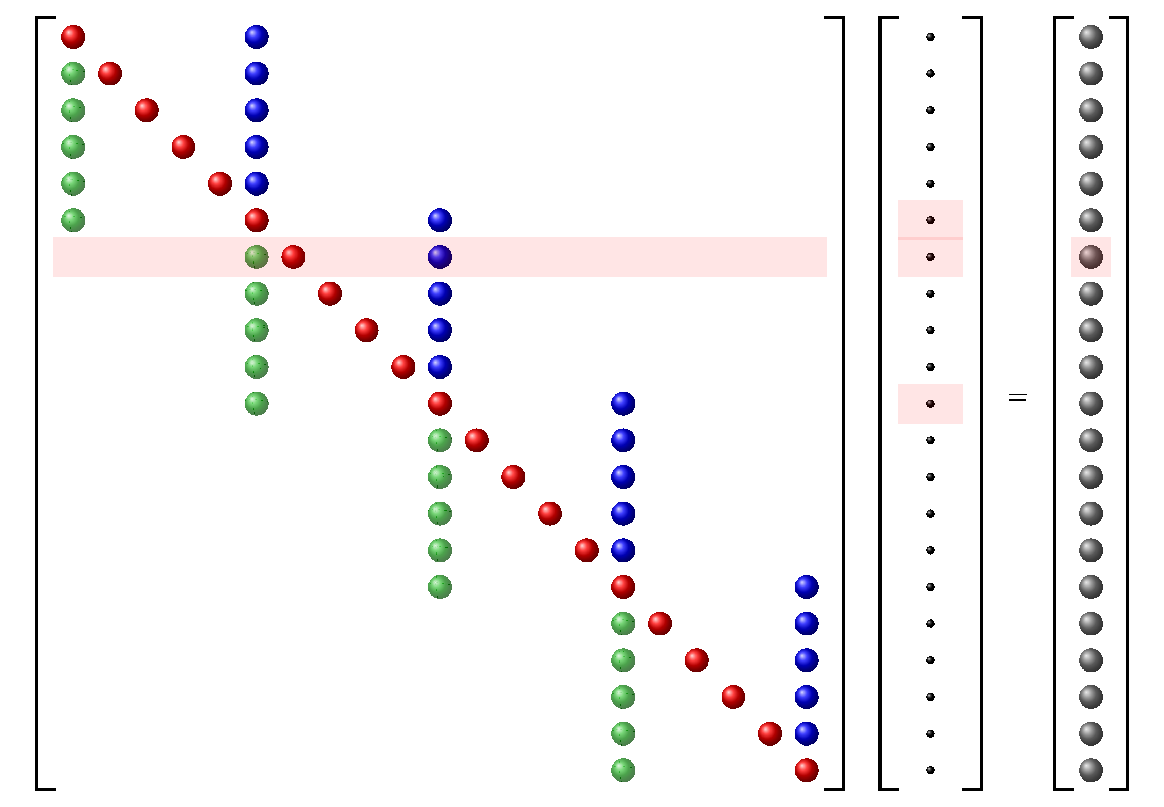} (f) \\  \hline
\end{tabular}
\end{centering} 
\caption{
\label{fig:method}
(Color online)
Schematic depiction of the proposed parallel reduction scheme.  For compactness,
all elements are labeled by colored bullets. 
(a) Schematic depiction of Eq.~(\ref{eq:BackwardStep}) and the partitioning scheme. 
Lower, center, and upper diagonals are described correspondingly by 
vectors $\mat{a}$, $\mat{b}$, and $\mat{c}$. Vector $\mat{d}$ is shown to the right. The vector
of unknowns $\mat{\Psi}$ is not shown since it is not used in the early stages of the algorithm.
The frames show ``blocks'' as described in the text. 
The lines between the framed regions are the joint lines $J_w$.
(b) The state of the matrix after the RD stage.
(c) Matrix after the RU stage.
(d) Matrix elements after the joint up and down stages, JU+JD.
Notice the characteristic stacked-``N'' shape.
The shape and stacking pattern is preserved if the blocks are different in size.
(e) The joint lines of the matrix from the previous stage form a reduced tridiagonal system of equations. The number
of elements equals to the number of parallel processes plus one.
This matrix is solved in the RS stage, producing the values
of $\mat{\Psi}$ on the joint lines. These are referred to as the support values.
(f) The non-joint lines of the reduced matrix
have the property that their elements can be solved independently by using the two surrounding support values. This property is
used during the SS stage. 
}

\end{figure*}

The entire algorithm is outlined in Table~\ref{tab:method}, and is described in a step-by-step manner below. A visual representation of the algorithm can be found in Fig.~\ref{fig:method}. 

\subsubsection{Partitioning of the matrix in blocks} \label{subsec:partitioning}
In our parallelization method, we start by deciding on the partitioning of the workload. This part does not involve computation. First, we select a number $n_{bl}$ of blocks (see Fig.~\ref{fig:method}.a),
and a corresponding number $n_{bl}+1$ of what we call \emph{joint  lines} of Eq.~(\ref{eq:BackwardStep}), which delimit the different blocks. 
The joint lines must include lines $1$ and $n_{tot}$, 
and a number of $n_{bl}-1$ lines within the matrix. We label the joint lines
as $J_w$, where the index of the joint lines $w$ (which is not altered during the whole propagation) runs from 0 to $n_{bl}$. Thus $J_0=1$ and $J_{n_{bl}}=n_{tot}$, and $J_w>J_{w-1}$. It is worth noting that such a construction allows blocks to differ in size.

We assume the program is executed on $n_{bl}$ parallel processes, which can be for instance MPI processes, or the threads of either a shared memory multiprocessor or a GPU.
We must point out that neither the tridiagonal matrix nor the complete wavefunction ever need to be placed entirely on a single process. Besides, the data can remain distributed throughout the entire execution and during  multiple time propagation steps, as will be explained below.

\subsubsection{Reduction stages RD and RU}
The algorithm starts with the downward (RD) and upward (RU) \emph{reduction} stages. During the reduction, each parallel process works on its block and performs operations similar to the Thomas method. There is no communication at this stage, and each parallel process accesses elements sequentially from its memory area.
In the new matrix generated after performing the RD (see Fig.~\ref{fig:method}.b) and RU (see Fig.~\ref{fig:method}.c) stages, the values stored in $\mat{b}$ and $\mat{c}$ no longer correspond to the lower and upper diagonals. Instead, 
elements $b_i$ are the values corresponding to column $J_w$ for $J_w<i<J_{w+1}$,
whereas elements $c_i$ contain values from column $J_{w+1}$.

\subsubsection{Joining stages JU and JD}
After the RD and RU reduction stages, we have to implement the upward (JU) 
and downward (JD) \emph{joining} stages (see Fig.~\ref{fig:method}.d), which involve communication.
In this step, each process writes to its surrounding joint lines.
Because two parallel processes write to the elements $a_{J_w}$ and $d_{J_w}$ on the joint lines,
an atomic operation is required. To distinguish such cases, in Table~\ref{tab:method},
we use the ``protected'' symbol $\atomicarrow$ for atomic assignment. 
Only four atomic operations per parallel process are required.
Notice that the writing thread does require the outcome of these atomic operations.
On systems with distributed memory, one may alternatively employ two extra memory elements per each joint line, and process the ``atomic'' additions afterward.
Computing the value of the factor $r_i$ for the joining stages JU and JD does not require communication. 

After the joining stage, the elements of the matrix are non-zero on the diagonal and in columns $J_w$, $w=0,\dots,n_{bl}$, and on rows $i$ with $J_{w-1} \le i\le J_{w+1}$. Thus, the matrix gets the characteristic form of stacked N-shapes, which 
gives the method its name. Here it is worth to mention that even though in the work
of Wang and Mou \cite{wangmou1991ieee} a block N-shaped matrix is also observed,
that matrix is intrinsically different in the relative placement of the N-shapes. This makes the scheme substantially different from the one presented here.

\subsubsection{Reduced solution RS stage and determination of the support values}
After finalizing the previous step, the joint lines represent a reduced tridiagonal matrix of size $n_{bl}+1$ (see Fig.~\ref{fig:method}.e). The solution of this system, which is implemented serially, is named \emph{reduced solution} (RS) stage which, as described in Table~\ref{tab:method}, can be implemented via Thomas reduction (more elaborated methods can be used as well). This step allows us to determine the $J_w$ components of $\mat{\Psi}$, which we name \emph{support values}.

\subsubsection{Support substitution SS stage and obtention of the wave function}
In the $w-$th block of the matrix after the RS stage, each line of index $J_{w-1}<i<J_w$, represents a linear equation (see Fig.~\ref{fig:method}.f) of the form 
\[
b_i \psi_{J_{w-1}} + a_i \psi_i + c_i \psi_{J_w} = d_i \;.
\] 
Since the support values have been determined in the previous stage, each of these equations can be solved independently for $\psi_i$. This is the final part of the algorithm, named \emph{support substitution} (SS) stage, where each block can perform independently and in parallel.

\begin{table}
\caption{
\label{tab:method}%
Stages of the N-shaped partition method, top-to-bottom.
Notice that stages JU and RU commute with each other.
Encircled left-value assignment denotes atomic addition.
For all parallel
processes, $w=0,\dots,n_{bl}-1$. The RS stage is the solution
of the reduced tridiagonal system. 
}

\begin{equation*}
  \setlength{\arraycolsep}{0pt}
  \renewcommand{\arraystretch}{1.2}
  \begin{array}{  l l @{\hspace{3px}} l  @{\hspace{6em}} r   @{\,} l  @{\,}  l   }
  \multirow{24}{*}{\rotatebox[origin=c]{90}{ \text{parallel} }}     
              \ldelim\{{24}{5mm}{ }
 &
  \multirow{6}{*}{\rotatebox[origin=c]{90}{  \text{RD stage} }  }     
              \ldelim[{6}{5mm}{ }
 & 
   \multicolumn{4}{l}{i=J_{w}+2,\dots, J_{w+1}-1 } \\
                                      &&& r_i   &\leftarrow & -b_{i}/a_{i-1}  \\
                                      &&& b_i &\leftarrow & r_i b_{i-1} \\
                                      &&& a_i &\leftarrow & a_i+r c_{i-1}\\
                                      &&& d_i &\leftarrow & d_i+r_i d_{i-1}\\
                                      &&& (c_i&\multicolumn{2}{l}{\text{ unchanged})}  \\
 &  \multirow{6}{*}{\rotatebox[origin=c]{90}{  RU\text{ stage} }  }     
              \ldelim[{6}{5mm}{ }
                                      
  &  \multicolumn{4}{l}{i=J_{w+1}-2,\dots, J_{w}+1}\\
                                      &&&   r_i &\leftarrow & -c_{i}/a_{i+1} \\
                                      &&& b_i &\leftarrow & b_i+ r_i b_{i+1} \\
                                      &&& c_i &\leftarrow & r_i c_{i+1}\\
                                      &&& d_i &\leftarrow & d_i+r_i d_{i+1}\\
                                      &&& (a_i & \multicolumn{2}{l}{\text{ unchanged})} \\
 & \multirow{6}{*}{\rotatebox[origin=c]{90}{  \text{JU stage} }  }     
              \ldelim[{6}{5mm}{ }
                                      
  &
     \multicolumn{4}{l}{i=J_{w}}\\ 
                                      &&&   r_i &\leftarrow& -c_{i}/a_{i+1} \\
                                      &&& c_i &\leftarrow& r_i c_{i+1}\\
                                      &&& a_i &\atomicarrow& a_i+r_i b_{i+1}\\
                                      &&& d_i &\atomicarrow& d_i+r_i d_{i+1}\\
                                      &&& (b_i& \multicolumn{2}{l}{\text{ unchanged})} \\
 & \multirow{6}{*}{\rotatebox[origin=c]{90}{  \text{JD stage} }  }     
              \ldelim[{6}{5mm}{ }                                      
  &  
     \multicolumn{4}{l}{ i=J_{w+1}}\\
                                      &&&    r_i &\leftarrow& -b_{i}/a_{i-1} \\
                                      &&&  b_i &\leftarrow& r_i b_{i-1} \\
                                      &&&  a_i &\atomicarrow& a_i+ r_i c_{i-1}\\
                                      &&&  d_i &\atomicarrow& d_i+r_i d_{i-1}\\
                                      &&&  (c_i& \multicolumn{2}{l}{\text{ unchanged})} \\ 
\\
 \multirow{8}{*}{\rotatebox[origin=c]{90}{ \text{serial  } }}     
              \ldelim\{{8}{5mm}{ }
 &
  \multirow{8}{*}{\rotatebox[origin=c]{90}{  \text{RS stage} }  }     
              \ldelim[{8}{5mm}{ }                                       
  &  
      \multicolumn{4}{l}{i=J_w,\,\text{ with }w=1,\dots,\,n_{bl} }\\ 
                                      &&&   r_i&\leftarrow& -b_{i}/a_{i-1} \\
                                      &&&  a_i&\leftarrow& a_i+r_i c_{i-1}\\ 
                                      &&&  d_i&\leftarrow& d_i+r_i d_{i-1}\\ 
  &&   
     \\ 
                                      &&&        \psi_{n_{tot}}&\leftarrow& d_{n_{tot}}/a_{n_{tot}} \\
  &&   
      \multicolumn{4}{l}{i=J_w,\,\text{ with }w=n_{bl}-1,\dots,0} \\ 
                                      &&&          \psi_i&\leftarrow& (d_i-c_i \psi_{i+1})/a_i \\
%
%
%
\\
 \multirow{3}{*}{\rotatebox[origin=c]{90}{ \text{parallel} }}     
              \ldelim\{{3}{5mm}{ }
 &
  \multirow{3}{*}{\rotatebox[origin=c]{90}{  \text{SS stage} }  }     
              \ldelim[{3}{5mm}{ }
 &   
      \multicolumn{4}{l}{i=J_{w}+1,\dots,J_{w+1}-1 } \\ 
                                       &&& r_i&\leftarrow& 1/a_i \\
                       &&&   \psi_i&\leftarrow&   ( d_i - b_i \psi_{J_{w}} - c_i \psi_{J_{w+1}} ) r_i  \\
%
%
%
\end{array}
\end{equation*}

\end{table}

A purely mathematical description of the algorithm can be found in Appendix I.

\subsubsection{Iterating}\label{sec:iter}

The reduced tridiagonal system obtained after 
the reduction and joining stages will often be rather small. 
Its size equals to the number of parallel processes used to solve the large system. For example, on a 10-core CPU one would need to solve an 11-element tridiagonal system.
In such cases, direct solution as shown in the RS stage in Table~\ref{tab:method} is fastest. However, one can foresee situations when the reduced system is rather large. For example, it is the case for the solution with a GPU. Since a GPU can support (and benefits from using) thousands of threads, the reduced system may be large as well. This large reduced tridiagonal system can be likewise solved with the same algorithm presented here. If we think of the algorithm as an operator $N$, acting on a given system $M_0$ (which comprises the tridiagonal matrix and the vectors $\mat{d}$ and $\mat{\Psi}$ as a whole), it can be schematically written as
\begin{equation*}
\begin{aligned}
    \text{N}[M_0]=&(\text{RD}[M_0]+\text{RU}[M_0])+(\text{JU}[M_0]+\text{JD}[M_0])\\
    &+\text{RS}[M_1]+\text{SS}[M_0] \;,
    \end{aligned}
\end{equation*}
where $M_1$ is the tridiagonal system at which we arrive after the JD stage. The plus sign is used as an aid to indicate the order in which the stages are performed. If we iterate the algorithm one time, we get
\begin{equation*}
\begin{aligned}
    \text{N}[M_0]=&(\text{RD}[M_0]+\text{RU}[M_0])+(\text{JU}[M_0]+\text{JD}[M_0])\\
    &+(\text{RD}[M_1]+\text{RU}[M_1])+(\text{JU}[M_1]+\text{JD}[M_1])\\
    &+\text{RS}[M_2]+\text{SS}[M_1]+\text{SS}[M_0]\;.
    \end{aligned}
\end{equation*}
Here we can see that solving RS[$M_1$] is equivalent to solving (RD[$M_1$]+RU[$M_1$])+(JU[$M_1$]+JD[$M_1$])+SS[$M_1$]. Furthermore we can see that by iterating, the serial RS stage is performed on the smaller system $M_2$, instead of $M_1$.
By induction, we can infer that the algorithm may be iterated a number $n_{it}$ of times in the following manner
\begin{equation}
\begin{aligned}
    \text{N}[M_0]=&\sum_{i=0}^{n_{it}}\left\{(\text{RD}[M_i]+\text{RU}[M_i])+(\text{JU}[M_i]+\text{JD}[M_i])\right\} \\
   & + \text{RS}[M_{n_{it}+1}]+\sum_{i=n_{it}}^0 \text{SS}[M_i] \;,
    \end{aligned}\label{eq:iteration}
\end{equation}
where we can see that by iterating the algorithm the portion that has to be performed serially is reduced and, at the same time, the size of the blocks (associated with the shared memory required in a practical implementation) is reduced as well. The only added costs may come as communication penalties which, as we will see later, can be masked. 

\subsection{Explicit forward half-step and incorporation into the Crank-Nicolson algorithm}\label{sec:explicit_forward_half_step}

While our scheme represents the implicit backward half-step of the Crank-Nicolson algorithm, it also accommodates naturally the explicit forward half-step. 
The explicit forward half-step is the computation of the r.h.s.\ of Eq.~(\ref{eq:ForwardStep}).
Since the matrix $\mat{T}^\dagger$ is also tridiagonal, the computation is straightforward. It is worth to notice that in this step no communication is required.

\section{Numerical performance: memory handling, communication masking, and efficiency}\label{sec:performance}

\subsection{Memory handling}

Our method involves not only a novel algorithm for parallelizing the computation of the Crank-Nicolson scheme, but also offers many advantages when the computing architectures are taken in consideration. As we describe in this section, communication masking, memory handling and efficiency are improved by our method. 

\subsubsection{Memory locality \label{sec:mem_locality}}
All data that relates to a given block, including the elements of the tridiagonal matrices and the wavefunction itself, can permanently reside on its dedicated parallel process.
This also applies between iterations of the algorithm where one wants to advance the wavefunction by multiple timesteps. With good design the same applies to multidimensional systems. The only data that needs to be gathered is on the joint lines. As explained above, the parallelization approach can be nested to apply to the joint lines as well. Thus, in principle, no system is too large to be propagated due to memory constraints on an individual parallel process, provided one has a sufficiently large number of such processes. 

Besides, on one hand, even though there will be communication penalties when the number of parallel processes is increased, part of this communication can be performed while solving the serial RS stage. On the other hand, one does not need to ever gather the wavefunction in a single location which, up to our knowledge, is unprecedented in parallel implementations of the Crank Nicolson scheme.

\subsubsection{Memory access patterns}

As mentioned above, the solver must access data in a sequential manner. It is only in this case that the values may be fetched fast enough to load the computing capacity of the processor. Our algorithm is designed to provide such an access pattern. All the loops on the parallel blocks access data sequentially, with the possible exception of the RS stage. However, the data for the (smallest) RS stage are located on the joint lines that are anyways separated in memory.

\subsection{Communication masking} \label{sec:communication}

\subsubsection{Communication and masking of the JD and JU stages}

The algorithm requires two synchronization points, before and after the RS stage.
For systems with distributed memory, the required communication involves collecting four values per process to perform the reduced solution. 
These are the values obtained on the joint lines after performing stages JU and JD.
After the RS stage, two support values $\mat{\Psi}$ must be made available to each process.
Thus the total communication cost is $6(n_{bl}+1)$ elements per step and does not depend on the original size of the matrix.

The communication cost can be furthermore reduced if one notices that the
JD stage can be performed already after RD. Thus, instead of performing the stages in the following order
\begin{equation*}
\text{(RD+RU)+(JU+JD)}\;,
\end{equation*}
one does it as
\begin{equation*}
\text{(RD+JD)+(RU+JU)}\;.
\end{equation*}
This also represents a better
memory access pattern. If $n_{tot}\gg n_{bl}$, it is plausible that the computing time of performing the RU and JU stages exceeds the network transport time of the data in the JD stage, and thus the JD communication time can be completely masked.

In the case when one is solving multiple tridiagonal systems at once,
most of the communication load is masked for as long as the 
processes are allowed to proceed to the next system while 
waiting for joining stage data from the previous system.

\subsubsection{Communication masking during the RS stage}
Some additional improvements are readily available which allow masking the cost of communication and that of performing the serial RS stage. While the RS stage is performed, perhaps by a single process, and while the resulting support values $\psi_{J_w}$ are being sent to other processes in the case of distributed-memory architecture, all other processes can work on computing values
of $1/a_i$ for the $r_i$ coefficients from the data located inside the blocks.
These operations do not require additional memory or communication. Under certain circumstances this calculation will mask the cost of the RS stage.

\subsubsection{Communication masking during the SS stage}
A generalization of the approach in the above section is pre-computing $1/a_i$ coefficients in the SS stage.
It does not require the completion of either of the communication stages, including the solution of the reduced system. Thus, upon sending the data from the JD stage, each parallel process can proceed to invert its diagonal elements. 
The cost of this operation may mask not only the communication but, importantly, the entire serial computation cost of the RS stage.

\subsubsection{Communication masking of a sequence of tridiagonal systems}
Suppose that several tridiagonal systems must be solved in a sequence, such as during the time evolution of a quantum system. The RD stage of the second tridiagonal system requires knowledge of the solution from the previous system. However, if one is willing to perform the storage of a column of the values of $r_i$ factors for the RD stages and possibly RU, then most of the operations for these two stages can be carried without the knowledge of the solution vector $\mat{d}$. This provides a powerful communication masking buffer because, with careful design, both the communication and the (serial) reduced stage can be masked by these calculations from the RD and RU stages. 
Once the information arrives for the solution of the reduced system (the support values $\psi_{J_w}$ and $\psi_{J_{w+1}}$ from the RS stage), the support values can be immediately used (as $\mat{d}$ values) in the RD and RU stages of the next system.

\subsection{Communication masking during the explicit forward half-step}
As we mentioned in Sec.~\ref{sec:explicit_forward_half_step}, the computation of the r.h.s.\ of Eq.~(\ref{eq:ForwardStep}) does not require communication. Furthermore, each parallel process computes its section 
of the vector $\mat{d}$, as determined by the partitioning indices $J_w$.
That is, each process $w$ computes elements $d_i$ for $J_w < i \le J_{w+1}$,
and additionally, the first process, $w=1$, computes the first element, $d_{J_0}=d_1$ (usually equal to zero). 

On distributed systems, the reduced system and its corresponding vector elements are likely to be serviced by a dedicated process. In such cases, each block computes its first and last $\mat{d}$ elements, sends them to the root, and continues to compute the rest of $\mat{d}$. Thus the communication here can also be masked. In fact, it is more efficient to incorporate the explicit forward half-step into the backward RD stage. In this way, the ``freshly'' computed values of
$d_i$ are immediately used in the reduction stage.

\subsection{Considerations for multi-dimensional problems}\label{sec:nd_considerations}
The method as described here applies directly only to 1D systems. 
As discussed in Section~\ref{sec:need_parallel}, multi-dimensional (ND) problems can sometimes be reduced to series of tri-diagonal problems. The nature of communication patterns depends on the chosen ND method and the representation of the wavefunction, which are domain-specific. 
Detailed discussion of this topic will be provided elsewhere. 

\subsection{Efficiency}

Within our scheme the total number of the computationally expensive divisions performed in parallel is $3(n_{tot}-n_{bl})$, while only $3n_{bl}$ divisions are performed serially. This compares favorably to the standard serial algorithm with its $3n_{tot}$ divisions. Furthermore, the $3n_{bl}$ serial divisions of our method can be further performed in parallel, as we describe in Sec.~\ref{sec:iter}, if the number of blocks is still large. On the other hand, the Woodbury formula exhibits a $9n_{tot}$ number of divisions which can be performed in parallel (which is comparable but larger than in our method) and a number $n_{bl}\times n_{tot}$ of multiplications and additions, while in our method we perform only $9n_{tot}$ multiplications and $6n_{tot}$ additions (see Appendix II).

\subsubsection{Performance estimation and optimization of a single- and multiple-iteration of the algorithm}
Our computation scheme reduces a tridiagonal system to identical but smaller systems.
The size of the smaller systems is not dependent on the size of the original system,
and therefore is fully controllable. Below we describe the general conditions for optimal execution of the algorithm and 
provide a theoretical estimation of the computing time. As mentioned in Sec.~\ref{sec:iter}, it is possible to recursively apply the method to the matrix obtained after the SS stage. The algorithm as described in Sec.~\ref{sec:method} represents what would be a single iteration. Here we will provide an estimation of the optimal block size and the theoretical performance for both a single- and multiple-iteration of the algorithm.

We begin with the case of a single iteration,
and assume the system has size $n_{tot}$ and is solved with $n_{bl}$ parallel processes.
Suppose the time to solve the RS stage (Fig.\ref{fig:method}.e)
is equal to $\beta \times n_{bl}$, where the coefficient $\beta$ depends on the architecture.
The size of parallel blocks is $n_{tot}/n_{bl}$, and the cumulative time for the RD, RU and SS stages can be described as $\alpha \times n_{tot}/n_{bl}$. Because the nature and quantity of the operations is the same as for the Thomas algorithm, we can expect the constants $\alpha$ and $\beta$ to be of the same order of magnitude. The stages JD and JU require a number of operations which demand a time proportional to $n_{bl}$, and involve communication. Nevertheless, as explained in detail in Sec.~\ref{sec:communication}, this communication can be masked and thus will be neglected in this analysis for simplicity.
Therefore the total execution time $t_{tot}$ can be modeled as 
\begin{equation*}
t_{tot} \sim  \alpha \frac{n_{tot}}{n_{bl}} + \beta n_{bl}\;.
\end{equation*}
The latter equation allows us to calculate, in an approximate way, the optimal number of blocks $n_{bl}^*$ which minimizes the execution time, 
\begin{equation}
n_{bl}^* \sim \delta \sqrt{n_{tot}}\;,
\label{eq:optimal_nblocks}
\end{equation}
where 
\begin{equation*}
\delta = \sqrt{\frac{\alpha}{\beta}}\;,
\label{eq:optimal_time}
\end{equation*}
is a system specific factor which will depend on the specific architecture employed. 
This tells us that the optimal number of parallel processes will also be proportional to $\sqrt{n_{tot}}$. Provided such resources are available, the overall execution time will likewise scale as $\sqrt{n_{tot}}$, a strong advantage over the Thomas algorithm that scales as $n_{tot}$.

For the case of several nested iterations, the execution time can be written as
\begin{equation*}
t_{tot,(n_{it})}\sim  \beta n_{bl,(n_{it})} +\alpha \sum_{i=0}^{n_{it}-1} \frac{n_{bl,(i)}}{n_{bl,(i+1)}}\;,
\end{equation*}
here we use the convention $n_{bl}^{(0)}=n_{tot}$.
We imply that some workers are made idle on subsequent steps, $n_{bl}^{(i)}\ge n_{bl}^{(i+1)}$.
In this case, the optimal configuration is again obtained by equally partitioning the work between all stages, i.e.
\begin{equation}n_{bl,(i)}^*/n_{bl,(i+1)}^*\sim n_{bl,(n_{it})}^*\sim n_{tot}^{1/(n_{it}+1)}\label{eq:optimal-P-multistage}\;.
\end{equation}
The total execution time is given by
\begin{equation}
t_{tot,(n_{it})}\sim (\alpha n_{it}+\beta) n_{tot}^\frac{1}{n_{it}+1}\;.
\label{ref:time-multistage}
\end{equation}

Finally, we can derive the theoretically optimal number of iterations for the limiting case of a massive parallel system 
\[
n_{it}^*\sim \log{n_{tot}},
\]
that leads to execution time that scales as
\[
t_{tot,(n_{it}^*)}\sim \alpha e \log(n_{tot})\;,
\]
which would represent a very advantageous scaling. It is worth to mention though, that in this work we did not test the minimum size of the system for which this hypothetical scaling is reached, which might be difficult to achieve in practice. Thus, the scaling in our work is expected to be governed by Eq.~\ref{ref:time-multistage}.

\section{Implementation on computing architectures} \label{sec:implementation}
\subsection{CUDA Implementation} \label{sec:cuda}

We have implemented the above method for execution on graphical processing units (GPU) produced by NVIDIA. The program was written with the CUDA-C programming language \cite{CUDA_6_ProgrammingGuide2014}. Optimally solving the tridiagonal systems on the GPU has received a great deal of attention %
\cite{Zhang2010-FastTridiagonalSolversOnTheGPU,Davidson2011,Hwu2011-AScalableTridiagonalSolverForGPUs,Chang2012,Goddeke2011}%
. For a review, see Ref.~\cite{WenHwu2014}. 
An efficient implementation must map well both to the GPU hardware and programming model.
The choice is usually made to explicitly load the partitions to the GPU shared memory by coalesced memory reads, and to solve (reduce, etc.) the loaded data block by all the threads involved in loading the data. As the number of data elements per thread is small, the choice naturally falls on the parallel cyclic reduction methods. We are instead using the GPU as a \emph{model} of a massively parallel system, treating global device memory as a cache-coherent memory space. The shared memory of the GPU streaming multiprocessors (SMX) was not directly addressed but was instead reconfigured to act as high-level cache 
\cite{CUDA_KeplerTuningGuide}. Each \emph{individual} thread was serving its own block. While one would in principle benefit from an interleaved matrix layout, we have decided to keep the conventional layout, and to rely on efficient memory cacheing. Since each block reads and writes most of its data sequentially, very good cacheing is to be expected. Indeed, we find that 77\% of all global loads hit the L1 cache (same as shared memory), 8\% hit the L2, and only 15\% have to be loaded from the global memory. This performance is not much below an explicit coalesced load to the shared memory. Likewise, the memory writes are almost always sequential, with 60\% stores hitting L2 cache. Overall, this provides sufficient memory bandwidth, given the added mathematical demand for complex arithmetics. This was the case despite the fact that the complex operations were written and optimized explicitly in terms of real and imaginary components, which significantly improved the throughput of complex arithmetic operations.
 
Global memory of the GPU is synchronized upon completion of every CUDA kernel. The staging of the algorithm fits the GPU model well. Separate kernels were written for RD+JD+RU+JU, RS, and SS stages. The communication corresponding to the JD and JU  stages was carried via atomic operation to the global memory. The kernels can be interleaved to solve the reduced system by the same parallel method, as per Eq.~(\ref{eq:iteration}). 

\subsubsection{Numerical results}

\begin{figure}
\begin{centering}
\includegraphics[width=.475\textwidth]{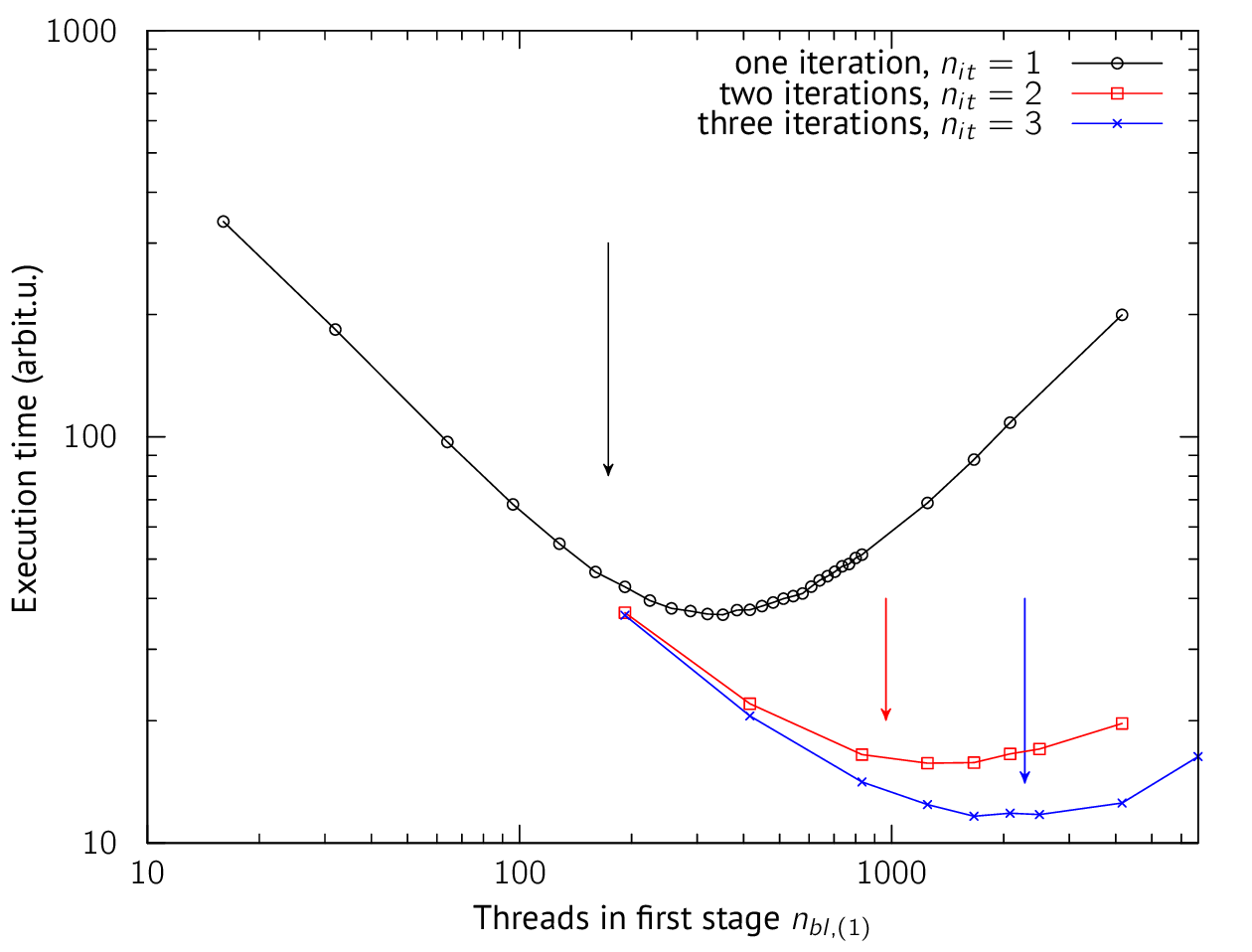}
\end{centering} 
\caption{
\label{fig:scaling-iterations-gpu}
Total execution time for computation with the GPU program on a box containing $n_{tot}=3\cdot 10^5$ spatial grid points. 
Shown are calculations with a one-iteration algorithm as outlined in Table~\ref{tab:method}, and nested-application multi-iteration methods in which the reduced system is also solved in parallel, as described in text. Arrows show the predicted (up to a fixed constant) optimal number of threads in the first iteration, according to Eq.~(\ref{eq:optimal-P-multistage}).
}
\end{figure}

In Fig.~\ref{fig:scaling-iterations-gpu} we show the total execution time for computation with the GPU program of tridiagonal matrices of size $n_{tot}=3\cdot 10^5$ spatial grid points. 
There we can see calculations with a one-iteration algorithm as outlined in Table~\ref{tab:method}, and nested-application multi-iteration methods in which the reduced system is also solved in parallel, as described in the text text. Arrows show the predicted (up to a fixed constant)  optimal number of threads in the first iteration, according to Eq.~(\ref{eq:optimal-P-multistage}). We can see a good agreement between the theoretical prediction and our numerical results, as well as the improvement of the computing performance when increasing the number of iterations. 

\subsection{MPI Implementation} \label{sec:mpi}

Apart from GPUs, our algorithm is suitable to be implemented in MPI, which allows for parallel execution on distributed systems. It is worth to mention that to achieve good performance the communications must be minimal and to a good degree masked by computation, there should not be excessive synchronization, and the data must be well parcelled to the memory of individual processes that ``own'' it. If such conditions are met, the algorithm, and even the same program, should be expected to perform at least as well on systems with cache-coherent (shared) memory, whether uniform or non-uniform access type. 

For the MPI implementation, it would be advantageous to assign the ``ownership'' of the joint lines to a separate process. This \emph{root process} could also be used to solve the reduced system. The remaining \emph{block processes} might maintain the elements corresponding to individual blocks. 
If running our implementation on several thousand cores, one can expect $n_{tot} \gg \sqrt{n_{tot}} \gg n_{bl} \gg 1$. 
That is, the number of elements served by the root process should be considerably smaller than the elements on each block process, as $n_{tot}/n_{bl} \gg n_{bl}$. For such a configuration, only one iteration would be used to solve the reduced system, and communications need only to be masked on the block processes.

\section{Conclusion}\label{sec:conclusion}

In this study we develop an algorithm for solving tridiagonal systems of linear equations. Our novel N-shaped partition method for parallel architectures greatly improves the performance compared to serial implementations. Furthermore, compared to other parallel computation schemes, our algorithm is architecture-agnostic and, compared to GPU-oriented methods, it can fully employ all the available memory and has even the potential to be practiced in mixed-architectures. 

In addition, we offer a mathematical characterization of the optimal block sizes. We also estimate the performance of this method which, depending on architecture factors, exhibits a square root dependence on the system size for a single iteration and approaches asymptotically a logarithmic dependence for a multiple nested iteration, compared to the linear scaling of the serial approach. We test numerically on a GPU the scaling of the efficiency of the method as a function of the number of times it is iterated. By this analysis we obtain good agreement with the mathematical prediction. This study allows us to show that our scheme can substantially improve the computation time even for a low number of nested iterations. To conclude our work, we also comment on the possible implementation of this method on MPI architectures.

Our contribution allows to perform otherwise prohibitive ab-initio calculations (because of both memory handling and computation time) for which large spatial grids are required like, for instance, the interaction of atoms and molecules with extremely long wave-length lasers, or first-principle calculations of the optical response of nanomaterials, among other possible applications.

\section*{Appendix I: Mathematical description of the algorithm}
In this section we will show our algorithm in mathematical form, leaving aside technical details of the implementation like communication, or specifying which part is done in parallel. This has to be understood as a formal mathematical representation of our proposed algorithm because in our scheme, for instance, we never sum two zeroes, nor perform sums that end up in a strictly zero element. 

Following the notation from Hoffman and Kunze's book~\cite{hoffman}, we will describe the elementary row operations of our scheme (Table~\ref{tab:method}) as a special type of function (rule) \textit{e} which associates each matrix \mat{T} with a matrix $e\left(\mat{T}\right)$. 
In the same way, we will define a function $f$, for operations on the vectors \mat{d}.

The matrix (row)operations on the RD and JD stages can be condensed into a single operation $e_d$ defined as
\begin{equation}   
e_{d}(\mat{T})_{l,j} = T_{l,j} \;\;\; \text{if  }l\neq i,\; e_{d}(\mat{T})_{l,j}=T_{i,j}+r_i T_{i-1,j} \;\;\; \text{if  }l=i\;,
\label{eq:ed}
 \end{equation}
where $r_i=-T_{i,i-1}/T_{i,i}$, while the vector operations can be characterized by an operation $f_d$ defined as
 \begin{equation}   
f_{d}(\mat{d})_{i} = d_i+r_i d_{i-1}\;.
\label{eq:fd}
 \end{equation}
These operations are applied from $i=J_w +2$ to $i=J_{w+1}$. 

Next, the matrix operations on the updated \mat{T} matrix on the RU and JU stages can be condensed into a single operation $e_u$ defined as
\begin{equation*}   
e_{u}(\mat{T})_{l,j} = T_{l,j} \;\;\; \text{if  }l\neq i,\; e_{u}(\mat{T})_{l,j}=T_{i,j}+r_i T_{i+1,j} \;\;\; \text{if  }l= i\;,
 \end{equation*}
where $r_i=-T_{i,i+1}/T_{i+1,i+1}$, while on the vector operations can be characterized by an operation $f_u$ defined as
 \begin{equation*}   
f_{u}(\mat{d})_{i} = d_i+r_i d_{i+1}\;.
 \end{equation*}
These operations are applied from $i=J_{w+1} -2$ to $i=J_{w}$. 

Next, for implementing the RS stage, we apply first the $e_d$ function (see Eq.~\ref{eq:ed}) on the matrix formed by the joint lines of the updated \mat{T} matrix, and $f_d$ (see Eq.~\ref{eq:fd}) on the elements corresponding to the joint lines of the updated \mat{d} vector. The index $i=J_w$  runs on the range $w=1,...,n_{bl}$ (joint lines). 

For the second part of the RS stage, we have to solve the algorithm backward to obtain the elements of the $\mat{\Psi}$ vector of unknowns. The $n_{tot}$ component of the vector is given by $\psi_{n_{tot}}=d_{n_{tot}}/a_{n_{tot}}$, while the rest of the components are obtained as
\begin{equation*}
    \psi_i=(d_i-c_i \psi_{i+1})/a_i \;,
\end{equation*}   
where index $i=J_w$ runs on the range $w=n_{bl}-1,...,0$ (joint lines).

Finally, we apply the SS stage to obtain the remaining components of the vector of unknowns
\begin{equation*}
    \psi_i=( d_i - b_i \psi_{J_{w}} - c_i \psi_{J_{w+1}} ) r_i\;, 
\end{equation*}
with $r_i=1/a_i$, and the index $i$ in the range $i=J_{w}+1,\dots,J_{w+1}-1$.

\section*{Appendix II: Comparison with the Woodbury formula}

Let us analyze how is the Woodbury formula \cite{numrec} applied to the parallel solution of a large tridiagonal system of equations. Being \textbf{T} an $n_{tot}\times n_{tot}$ tridiagonal matrix represented as
\begin{equation*}
\textbf{T}=
\begin{pmatrix}
  \begin{matrix}
a_{(1)} & c_{(1)} &  \cdots&  \cdots & 0 \\
b_{(2)} & a_{(2)} &  \cdots & \cdots& \cdots \\
\cdots & \cdots &  \cdots & \cdots& \cdots \\
\cdots  & \cdots  &  \cdots & a_{(n_{tot}-1)} & c_{(n_{tot}-1)}  \\
0 & \cdots & \cdots&  b_{(n_{tot})} & a_{(n_{tot})}
  \end{matrix}
\end{pmatrix} \;,
\end{equation*}
we want to obtain the vector $\bf{\Psi}$ which is the solution of 
\begin{equation*}
    \textbf{T}\bf{\Psi}=\textbf{d}\;,
\end{equation*}
for a given vector \textbf{d}. For that, we rewrite the matrix as
\begin{equation*}
    \textbf{T}=\textbf{T}_0+\textbf{C} \;,   
        \label{eq:defs}
\end{equation*}
where $\textbf{T}_0$ is  a block diagonal matrix consisting of $n_{bl}$ tridiagonal blocks
, and \textbf{C} is an $n_{tot}\times n_{tot}$ tridiagonal matrix of rank 2$n_{bl}$. 
We will assume for simplicity that the matrix is divided into blocks of equal size $n_{tot}/n_{bl}$, even though blocks of different sizes can be chosen. 
The correction matrix \textbf{C} is given by the product 
\begin{equation}
    \textbf{C}=\textbf{U} \textbf{V}^T    \;\;\;,
\end{equation}
where $\textbf{U}$ and $\textbf{V}$ are $n_{tot}\times n_{bl}$ matrices composed of a number $n_{bl}$ of column vectors $\textbf{u}_w$ and $\textbf{v}_w$ of size $n_{tot}$, each of which consists of, at most, two non-zero elements at the $J_w$ and $J_{w}+1$ components.  They can be represented as
\begin{equation}
\textbf{u}_w=
\begin{pmatrix}
  \begin{matrix}
\cdots \\
\cdots  \\
0  \\
u_{(J_w)}
  \end{matrix}
  \\
    \hline
\begin{matrix}
u_{(J_w+1)} \\
0  \\
\cdots  \\
\cdots
  \end{matrix}
\end{pmatrix}
\;\;\;,\;\;\;
\textbf{v}_w=
\begin{pmatrix}
  \begin{matrix}
\cdots \\
\cdots  \\
0  \\
v_{(J_w)}
  \end{matrix}
  \\
    \hline
\begin{matrix}
v_{(J_w+1)} \\
0  \\
\cdots  \\
\cdots
  \end{matrix}
\end{pmatrix} \;.
\label{eq:defvecuv}
\end{equation}
The non-zero elements of these vectors can be defined as
\begin{equation}
    \begin{aligned}
        u_{(J_w)}&=-a_{(J_w)}\;\;,\;\;u_{(J_w+1)}=b_{(J_w+1)}\;\;,\;\;\\
        v_{(J_w)}&=1\;\;,\;\;v_{(J_w+1)}=-c_{(J_w)}/a_{(J_w)}\;.   
    \end{aligned}
      \label{eq:uvcoeff}
\end{equation}
From these equations, we can see that the definition of the matrix $\textbf{V}$ requires $n_{bl}$ divisions.

The block diagonal matrix $\textbf{T}_0$ can be represented as:

\begin{strip}
\begin{equation*}
\textbf{T}_0=
\begin{pmatrix}
\begin{matrix}
 \ddots & \cdots & \cdots \\
 \cdots  & a_{(J_w-1)} &c_{(J_w-1)}  \\
\cdots & b_{(J_w)} & a'_{(J_w)} 
\end{matrix}
  & \rvline & \bigzero & \rvline & \bigzero  \\
\hline
  \bigzero & \rvline &
  \begin{matrix}
a'_{(J_w+1)} & c_{(J_w+1)} & \cdots&  \cdots & 0 \\
b_{(J_w+2)} & a_{(J_w+2)} &   \cdots& \cdots& \cdots \\
\cdots & \cdots & \cdots &  \cdots&  \cdots \\
\cdots  & \cdots   & \cdots& a_{(J_{w+1}-1)} & c_{(J_{w+1}-1)}  \\
0 & \cdots & \cdots &  b_{(J_{w+1})} & a'_{(J_{w+1})}
  \end{matrix}
 & \rvline & \bigzero \\
\hline
 \bigzero & \rvline   & \bigzero  & \rvline &
  \begin{matrix}
a'_{(J_{w+1}+1)} & c_{(J_{w+1}+1)} & \cdots  \\
b_{(J_{w+1}+2)} & a_{(J_{w+1}+2)} & \cdots  \\
\cdots  & \cdots  & \ddots 
  \end{matrix}
\end{pmatrix}\;.
\end{equation*}
\end{strip}

The modified diagonal elements of this matrix  (compared to the original matrix $\textbf{T}$), which are two elements near each joint line, are given by
\begin{equation}
    a'_{(J_w)}=2a_{(J_w)}\;\;,\;\;a'_{(J_w+1)}=a_{(J_w+1)}-b_{(J_w+1)}c_{(J_w)}/a_{(J_w)} \;.
\end{equation}
Because the division involved in obtaining these elements is already performed for setting up $\textbf{V}$ (see Eq.~\ref{eq:uvcoeff}), it is only required to execute $2n_{bl}$ multiplications and $n_{bl}$ additions for their definition.

Once these matrices have been defined, the inverse of $\textbf{T}$ can be obtained from
\begin{equation*}
    \textbf{T}^{-1}=\textbf{T}_0^{-1}-\left[\textbf{T}_0^{-1}\textbf{U} \left(\textbf{1}+ \textbf{V}^T \textbf{T}_0^{-1} \textbf{U}\right)^{-1}\textbf{V}^T\textbf{T}_0^{-1}\right]\;.
\end{equation*}
This formula is not used directly though. To get the solution vector $\bf{\Psi}$, we need to perform the intermediate steps we describe below. 

First, we have to solve the system of equations (remember that \textbf{U} is a matrix)
\begin{equation}
    \textbf{T}_0 \textbf{Z}= \textbf{U}   \;\;\;,
        \label{eq:s0uz2}
\end{equation}
from  which we get a matrix $\textbf{Z}$ of size $n_{tot}\times n_{bl}$, with $2n_{tot}$ non-zero elements. Here we have to solve the Thomas algorithm $2n_{bl}$-times for pairs of tridiagonal blocks (see Eq.~\ref{eq:defvecuv}). Therefore, the total complexity of this step is $6 n_{tot}$ divisions, and $8 n_{tot}$ multiplications and additions. 
 
Next, we have to solve the following system of equations for a newly defined vector $\textbf{y}$
\begin{equation}
    \textbf{T}_0 \textbf{y} = \textbf{d} \;\;\;.
        \label{eq:s0b}
\end{equation}
This implies solving the Thomas algorithm in each of the $n_{bl}$ blocks, with a total number of $3 n_{tot}$ divisions, and $4 n_{tot}$ multiplications and additions. 

After this step, we have to construct the following matrix
\begin{equation}
    \textbf{W}=\left( \textbf{1}+\textbf{V}^T \textbf{Z} \right) \;\;\;.
        \label{eq:defh}
\end{equation}
Here $\textbf{W}$ is an $n_{bl}\times n_{bl}$ tridiagonal matrix. It is possible to define this matrix by performing only $2n_{bl}$ multiplications and $3n_{bl}$ additions. With this matrix, we will solve the following system of equations to obtain an $n_{bl}$-size vector $\textbf{r}$
\begin{equation}
    \textbf{W}\textbf{r}=\textbf{V}^T \textbf{y} \;.
    \label{eq:solvew}
\end{equation}
Here we have to perform first $2 n_{bl}$ multiplications and additions for obtaining the right-hand side, and then apply the Thomas algorithm which will imply executing $3n_{bl}$ divisions, and $4n_{bl}$ multiplications and additions.

Finally, we get the solution from
\begin{equation}
    \bf{\Psi}=\textbf{y}-\textbf{Z}\textbf{r} \;\;\;,
        \label{eq:defx}
\end{equation}
for which it is necessary to perform $n_{tot}\times n_{bl}$ multiplications and additions to get the product $\textbf{Z}\textbf{r}$ and finally $n_{tot}$ additions.

We can see that the complexity of the method is mainly characterized by its $9 n_{tot}$ divisions (the execution of which can be parallelized), and its $n_{tot}\times n_{bl}$ multiplications and additions. 

When considering the iteration of the method, one way to implement it is to define a large number of blocks $n_{bl}$, which makes Eqs.~\ref{eq:s0uz2} and \ref{eq:s0b} highly parallelizable. Later one could then again partition the matrix $\textbf{W}$ (Eq.~\ref{eq:defh}), thus iterating the method in Eq.~\ref{eq:solvew}. We have to consider though that then, we have to perform a large number ($n_{tot}\times n_{bl}$) of multiplications and additions. Nevertheless, smarter implementations might be developed, which go beyond the scope of our work.

\section*{Acknowledgement}
This research and publication were supported by the Deutsche Forschungsgemeinschaft (DFG, German Research Foundation), 
SFB 1477 "Light-Matter Interactions at Interfaces", project number 441234705.

\bibliographystyle{elsarticle-num}
\bibliography{crank-nicolson}

\end{document}